\documentclass[11pt]{article}
\usepackage{a4wide}
\usepackage{hyperref}

\newcommand{\be}{\begin{equation}}
\newcommand{\ee}{\end{equation}}
\newcommand{\ba}{\begin{eqnarray}}
\newcommand{\ea}{\end{eqnarray}}
\newcommand{\non}{\nonumber}
\newcommand{\lang}{\left\langle}
\newcommand{\rang}{\right\rangle}
\newcommand{\al}{&\!\!\!\!}

\begin{document}
\title{ \hfill{\scriptsize SLAC-PUB-14347, FZJ-IKP-TH-2011-01} \\[1.5em]
       Isospin splittings of doubly heavy baryons}
\author{Stanley J. Brodsky,$^1$\footnote{{\it E-mail address:} sjbth@slac.stanford.edu} %
       ~Feng-Kun Guo,$^2$\footnote{{\it E-mail address:} fkguo@hiskp.uni-bonn.de} %
       ~Christoph Hanhart,$^{3}$\footnote{{\it E-mail address:} c.hanhart@fz-juelich.de} %
       ~Ulf-G. Mei{\ss}ner$^{2,3}$\footnote{{\it E-mail address:} meissner@hiskp.uni-bonn.de}\\[2mm]%
             \it \small $\rm ^1$SLAC National Accelerator Laboratory, Stanford University,
             Stanford, CA 94309, USA\\
             \it \small $\rm ^2$Helmholtz-Institut f\"ur Strahlen- und
             Kernphysik and Bethe Center for Theoretical Physics,\\
             \it \small Universit\"at Bonn,  D--53115 Bonn, Germany\\
             \it \small $\rm ^3$Institut f\"{u}r Kernphysik, J\"ulich Center
             for Hadron Physics and Institute for Advanced Simulation,\\
             \it \small  Forschungszentrum J\"{u}lich, D--52425 J\"{u}lich, Germany}%

\date{\today}

\maketitle

\begin{abstract}
The SELEX Collaboration has reported a very large isospin splitting of doubly
charmed baryons. We show that this effect would imply that the doubly charmed
baryons are very compact. One intriguing possibility is that such baryons have a
linear geometry $Q-q-Q$  where the light quark $q$ oscillates between the two
heavy quarks $Q$, analogous to a linear molecule such as carbon dioxide.
However, using conventional arguments,  the size of a heavy-light hadron is
expected to be around $0.5$~fm, much larger than the size needed to explain the
observed large isospin splitting. Assuming the distance between two heavy quarks
is much smaller than that between the light quark and a heavy one, the doubly
heavy baryons are related to the heavy mesons via heavy quark--diquark symmetry.
Based on this symmetry, we predict the isospin splittings for doubly heavy
baryons including $\Xi_{cc},\Xi_{bb}$ and $\Xi_{bc}$. The prediction for the
$\Xi_{cc}$ is much smaller than the SELEX value. On the other hand,  the
$\Xi_{bb}$ baryons are predicted to have an isospin splitting as large as
$(6.3\pm1.7)$~MeV. An experimental study of doubly bottomed baryons is therefore
very important to better understand the structure of baryons with heavy quarks.
\end{abstract}


\thispagestyle{empty}

\newpage

\section{Introduction}

A key prediction of QCD is the existence of baryons with two or three charm or
bottom quarks. Several years ago, evidence for the hadroproduction of five
different baryons with two charm quarks was reported by the SELEX Collaboration
at
Fermilab~\cite{Mattson:2002vu,Moinester:2002uw,SELEX,Ocherashvili:2004hi,Engelfried:2005kd,Engelfried:2007at}.
Two singly charged states $\Xi_{cc}^+(3443)$ and $\Xi_{cc}^+(3520)$ were
observed in the $\Lambda_c^+K^-\pi^+$ mass distribution, and three doubly
charged states $\Xi_{cc}^{++}(3460)$, $\Xi_{cc}^{++}(3541)$ and
$\Xi_{cc}^{++}(3780)$ were observed decaying into
$\Lambda_c^+K^-\pi^+\pi^+$~\cite{Mattson:2002vu,Moinester:2002uw,SELEX}. The
$\Xi_{cc}^+(3520)$ was also observed in the $pD^+K^-$~\cite{Ocherashvili:2004hi}
and $\Xi_c^+\pi^+\pi^-$ final states~\cite{Engelfried:2007at}.  An analysis of
the helicity angular distribution support the assignments that the
$\Xi_{cc}^+(3443)$ and $\Xi_{cc}^{++}(3460)$ form an isospin doublet, and the
$\Xi_{cc}^+(3520)$ and $\Xi_{cc}^{++}(3541)$ form another.  The preliminary
isospin mass splittings were reported to be 17~MeV and 21~MeV,
respectively~\cite{SELEX}. This observation is very puzzling because such values
are much larger than all isospin mass splitting of hadrons known so far. For
instance, the mass difference between the proton and neutron is
$m_n-m_p=1.29$~MeV, and between the charged and neutral $D$ mesons is
$M_{D^+}-M_{D^0}=4.77\pm0.10$~MeV~\cite{PDG2010}. The largest isospin splitting
ever observed is the double-strange baryons
$M_{\Xi^-}-M_{\Xi^0}=6.85\pm0.21$~MeV~\cite{PDG2010}.

Recently, the
mass of the lowest $\Xi_{cc}^{++}$ state was updated from  3460 to 3452~MeV~\cite{SELEXupdate}.
Although the isospin splitting is decreased from 17~MeV to 9~MeV for the
lower doubly charmed baryon isospin-doublet, it is still larger than all the
other known isospin splittings.
It is thus interesting to see whether it
is possible to obtain the observed rather large values from known physical
principles with controlled uncertainty. In Section~\ref{sec:implication}, it
will be shown that the SELEX observations would imply the $\Xi_{cc}$ to be very
compact, which, however, cannot be understood by any known mechanism of the
strong interactions.

Predictions for the isospin splittings of doubly heavy baryons will be presented
in Section~\ref{sec:QQq} based on the conventional assumption that the two heavy
quarks constitute a compact diquark. We then can apply the ansatz of heavy
quark--diquark symmetry.  Our predictions for the doubly charm baryons are
similar in magnitude to the isospin splittings for other hadron isospin
multiplets, but considerably smaller than the SELEX data. We will also give
predictions for the isospin splittings of the $cccc\bar q$ and $bbbb\bar q$
pentaquark states based on a heavy quark--``quadra-quark'' symmetry in
Section~\ref{sec:pentaquark}.

SELEX used $p_{\rm lab}=600$ GeV/c  $\pi^-, \Sigma^-$, and proton beams on a
nuclear target to produce the doubly charm baryons. A striking feature of the
SELEX measurements is the fact that the observed doubly charmed baryons are all
produced at  $x_F > 0.1$ (SELEX only has sensitivity in that region); i.e., at a
significant fraction of the projectile momentum. This is consistent with the ISR
measurements of the $\Lambda_c$~\cite{Bari:1991in} and the
$\Lambda_b$~\cite{Bari:1991ty} at high $x_F$, as well as   NA3 measurements at
CERN~\cite{Badier:1982ae} which showed that two $J/\psi$'s are hadro-produced at
high $x_F$ in pion-nucleus collisions; in fact, each $\pi A \to J/\psi J/\psi X$
event measured by NA3 has four charmed quarks with a flat  longitudinal momentum
distribution for $x_F > 0.4.$~\cite{Vogt:1995tf}. The SELEX and NA3 measurements
cannot be explained if the heavy quarks only arise from gluon splitting;
however, this is a natural consequence of the existence of intrinsic heavy
quarks in the
projectile~\cite{Brodsky:1980pb,Brodsky:1984nx,Harris:1995jx,Franz:2000ee}, such
as the rare $|uud c \bar c c \bar c\rangle$ Fock state in the proton or the
$|\bar u d c \bar c c \bar c\rangle$ Fock state  in the $\pi^-$. Since the
momentum distribution in such Fock states is maximized at low invariant mass,
all of the quarks tend to have the same rapidity and small transverse momentum.
The heavy quarks have the maximum momentum fractions in such configurations
since equal rapidity implies $x_i \propto \sqrt{m^2_i+ k^2_{\perp i}}.$  The
doubly charmed $ccq$ baryons are then formed in a collision by the
coalescence~\cite{Brodsky:2007yz,Brodsky:2006wb} of the comoving heavy quarks
with a light quark of the projectile --- the domain where the wave function of
the produced doubly charm hadron is maximal.  This mechanism also explains why
doubly charmed baryons are not readily produced in $e^+ e^-$ annihilation;  in
that case it is rare for the two charmed quarks to be in the same kinematic
domain.  The intrinsic charm mechanism also accounts for the non-factorized
nuclear-target dependence~\cite{Hoyer:1990us,Leitch:2004kr} of $J/\psi$
hadroproduction~\cite{Brodsky:2007yz,Brodsky:2006wb}. It also points to the high
$x_F$ domain of hadroproduction as the best kinematic region to search for heavy
hadron systems in general. Thus the best opportunity to create superheavy
hadrons and test their properties is in hadron-hadron collisions at high $x_F$
using the intrinsic heavy quark Fock state mechanism --- for example, at the
LHCb, or at future fixed-target experiments using the $7$ TeV  LHC beam.

The last section
contains a brief summary of our results.

\section{Implication of a large isospin splitting}
\label{sec:implication}

It is instructive to ask the question: What does a large isospin splitting imply
for the doubly heavy baryons? Isospin splittings originate from two sources ---
the $u$ and $d$ quark mass difference as well as electromagnetic contributions.
The interference pattern of the two different contributions to the mass
differences can be easily understood. The repulsive (attractive) Coulomb
interaction gives positive (negative) contribution to the electromagnetic (em)
self-energy of the baryons, so that the baryon with more absolute electric
charge has more em self-energy. The sign of the quark mass difference
contribution reflects the fact that the down quark is heavier than the up quark.
Hence, for the $\Xi_{cc}$ and $\Xi_{bc}$, the interference is destructive while
for the $\Xi_{bb}$ it is constructive. The sign of the em contribution to
$M_{\Xi_{cc}^{++}}-M_{\Xi_{cc}^{+}}$ is the same as the SELEX data, however, the
em and the quark mass term interfere destructively substantially reducing the em
effect. Thus, in order to quantitatively understand the SELEX data an unusually
large em contribution to the mass differences is necessary.

In the case of heavy particles the em effect is Coulombic since the magnetic
contribution is negligible. To quantify the em contribution, one may employ the
Cottingham formula to analyze the contribution of virtual photons. It can be
used to relate the em self-energy to the em form factor of a particle, see
e.g.~\cite{Gasser:1982ap}. Then the em self-energy is given
by~\cite{Gasser:1982ap} (neglecting the inelastic contributions)
\ba%
M^{\rm em} = \frac{\alpha Q^2}{4\pi^2}\int
\frac{d^3q}{\mathbf{q}^2} [G_E(-\mathbf{q}^2)]^2,
\ea %
where $\alpha=1/137.06$ is the fine structure constant, $Q$ is the total
electric charge in units of the proton charge, and $G_E(t)$ is the Fourier
transform of the charge distribution of the particle.  Taking a dipole
distribution with a mass parameter $m$ in units of GeV (which is sufficient  for
our purpose)
\ba %
\label{eq:emff} G_E(t) = \frac1{(1-t/m^2)^2},
\ea %
one can
perform the integration analytically, and gets
\ba %
M^{\rm em} = \frac{5}{32}\alpha Q^2 m.
\ea %
For a first estimate, let us consider the $ccq$ as
a two-particle system with charge $4/3$ and $e_q$. Therefore, the em
contribution to the isospin splitting of the doubly charmed baryons
$\delta_{\Xi_{cc}}\equiv M_{\Xi_{cc}^{++}}-M_{\Xi_{cc}^{+}}$ with $\Delta Q^2=
3$ is
\ba %
\label{eq:CotXicc} \delta_{\Xi_{cc}}^{\rm em} = \frac{15}{32}\alpha m
= 0.0034~m.
\ea %
The mean square radius of a particle is obtained by taking the
first derivative of its em form factor. For a heavy particle with the em form
factor given by Eq.~(\ref{eq:emff}), it is
\ba %
\langle r^2\rangle = \left.6 \frac{dG_E(t)}{dt}\right\vert_{t=0} =
\frac{12}{m^2}. \ea

Since the $\Xi_{cc}^{++}$ and $\Xi_{cc}^{+}$ contain $ccu$ and $ccd$ quarks,
respectively, the contribution to $\delta_{\Xi_{cc}}$ from the quark mass
difference is negative, i.e. $\delta_{\Xi_{cc}}^{\rm strong}<0$ because the $u$
quark is lighter than the $d$ quark. Hence, the em contribution must be larger
than the total isospin splitting of the doubly charmed baryons,
$\delta_{\Xi_{cc}}^{\rm em}> \delta_{\Xi_{cc}}$. If we take $9$~MeV as the
isospin splitting, one gets $m>2.65$~GeV from Eq.~(\ref{eq:CotXicc}), and
\ba%
\sqrt{\lang r^2\rang} < 0.26~{\rm fm}.
\ea%
This value is much smaller than the typical size of a hadron containing light
quark(s). In fact, it
is of similar size as the distance between the two heavy quarks $\sim 1/(m_Qv)$,
with $m_Q$ and $v$ the mass and the velocity of the heavy quark, respectively.
If we use a larger isospin splitting, e.g. 17~MeV, instead, the resulting
$\sqrt{\lang r^2\rang} < 0.14~{\rm fm}$ is even smaller.

Therefore, we conclude that a large isospin splitting would imply the doubly
heavy baryon to be very compact --- the larger the splitting, the smaller the size.
This conclusion can be easily understood because a large isospin splitting would
mean a large em self-energy which, being proportional to $\lang1/r\rang$, in
turn would mean a small size of the doubly heavy baryon.

One intriguing possibility is that doubly charm baryon states have a linear
geometry $Q-q-Q$ where the light quark $q$ oscillates between the two heavy
quarks $Q$, analogous to a linear molecule such as carbon dioxide ${C O_2} =
O-C-O$. In this case the overall size of the baryon would be relatively small. A
lattice gauge theory investigation of this possibility would be interesting.
However,  we are not aware of a mechanism in quantum chromodynamics (QCD) which
can keep the light quark in line and close to the heavy quarks. In particular,
if we take a Coulomb plus linear potential as the interquark interaction, the
distance between the light quark and a heavy quark scales as $(\sigma m_{\rm
cons})^{-1/3}\sim 1/\Lambda_{\rm QCD}$ with the flavor-independent string
tension $\sigma$ being the coefficient in front of the linear potential and
$m_{\rm cons}$ the constituent light quark mass. The numerical value for
$\sqrt{\sigma}$ is about 430~MeV from Regge trajectories of light mesons and
also from heavy quarkonia spectrum, see e.g.~\cite{Bali:2000gf}.  Thus the size
of a heavy-light hadron is expected to be around $0.5$~fm, much larger than the
size needed to explain the observed large isospin splitting.

\section{Isospin splittings of doubly heavy baryons -- a quantitative analysis}
\label{sec:QQq}

In the following we will assume that the distance between the light quark and a
heavy quark is much larger than the distance between the two heavy quarks inside
the doubly heavy baryon. Based on this conventional assumption, it was proposed
that there is a heavy quark--diquark symmetry~\cite{Savage:1990di} which
relates a doubly heavy baryon containing two heavy quarks to a heavy meson
containing a heavy anti-quark.

The distance between the two heavy quarks in a doubly heavy baryon is
characterized by $r_{QQ}\sim 1/(m_Q v)$, with $v$ being the heavy quark velocity.
It is much larger than the length scale for the light quark
$r_{qQ}\sim 1/\Lambda_{\rm QCD}$, as already discussed in the previous section.
Hence, one may perform an expansion in $r_{QQ}/r_{qQ}$ with a controlled
uncertainty. In the heavy quark limit, only the leading term is necessary, which
means the diquark formed by the two heavy quarks is point-like. Furthermore, the
diquark has the same color charge as a heavy anti-quark. Hence, there is an
approximate U(5) symmetry relating the spin-1/2 heavy anti-quark and the spin-1
heavy diquark, called heavy quark-diquark symmetry~\cite{Savage:1990di}. Using
this symmetry, the doubly heavy baryons can be studied by relating them to the heavy
mesons. In this section, we will calculate the em as well as quark mass
difference contribution to the isospin splitting of doubly heavy baryon masses
at next-to-leading order (NLO) in the chiral expansion. This is the lowest order at which isospin
breaking operators appear. In view of the present knowledge of the masses of the doubly
heavy baryons, this should suffice. Both effects can be taken into account
systematically up to a given order using  chiral perturbation theory with
virtual photons~\cite{Ecker:1988te,Urech:1994hd}. We will use the formalism
proposed in Ref.~\cite{Hu:2005gf}, which combines heavy mesons and doubly heavy
baryons into the same field, and construct the NLO chiral Lagrangian which is
responsible for the isospin mass splittings.

The super-flavor multiplet of heavy mesons and doubly heavy baryons can be
collected into a single $5\times2$ matrix field~\cite{Hu:2005gf}, which can be
written in components as
\ba%
\mathcal{H}_{a,\mu\beta} = H_{a,\alpha\beta} + T_{a,i\beta}
\ea%
where $a=u,d,s$ is the light flavor index, $\mu$ runs from 1 to 5,
$\alpha,\beta=1,2$ and $i=1,2,3$. The fields for the heavy mesons, $H_a$, and and
doubly heavy baryons, $T_a$, are given by
\ba%
H_{a,\alpha\beta} \al=\al \vec{P}_a^* \cdot \vec{\sigma}_{\alpha\beta} +
P_a\delta_{\alpha\beta}, \non\\
T_{a,i\beta} \al=\al \sqrt{2} \left(\Xi^*_{a,i\beta} +
\frac1{\sqrt{3}}\Xi_{a,\alpha}\sigma^i_{\alpha\beta}\right),
\ea%
where $P^{(*)}_a$ and $\Xi_{a}^{(*)}$ are the fields annihilating the vector
(pseudoscalar) heavy mesons and the spin-1/2 (3/2) doubly heavy baryons,
respectively. The field for the spin-3/2 baryon is constrained by
$\Xi^*_{a,i\beta}\sigma^i_{\beta\gamma}=0$.

One can construct the effective Lagrangian for the mass terms assuming the heavy
quark--diquark symmetry. At leading order, there is no  isospin splitting within
the same multiplet as can been from the Lagrangian constructed in
Ref.~\cite{Hu:2005gf}. At NLO, the Lagrangian relevant for the isospin and SU(3)
mass differences reads
\ba%
\label{eq:Lmass} \mathcal{L}_{\rm ISV} \al=\al -c {\rm Tr}
\left[\mathcal{H}_a^\dag \mathcal{H}_b^{} (\chi_+)_{ba}\right] - d_0 {\rm Tr}
\left[\mathcal{H}_a^\dag \hat{q}\mathcal{H}_b^{} (Q_+)_{ba}\right] \non\\\al\al
- {\rm Tr} \left\{\mathcal{H}_a^\dag \mathcal{H}_b^{} \left[d_1
(Q_+^2-Q_-^2)_{ba}+d_2(Q_+\langle Q_+\rangle)_{ba} \right] \right\},
\ea%
where the heavy-flavor charge operator $\hat{q}$ is defined as $\hat{q} H_a =
q_Q H_a$ and $\hat{q} T_a = -2q_Q T_a$, with $q_Q$ being the charge of the heavy
quark in a heavy meson. It is similar to the isospin breaking terms in the
Lagrangians constructed for heavy mesons~\cite{Guo:2008gp} and singly heavy
baryons~\cite{Guo:2008ns}. The operators $\chi_+$ and $Q_{\pm}$ contain the
Goldstone boson fields which are needed for higher order calculations
\ba%
\chi_+ = u^\dagger \chi u^\dagger + u\chi u ,\qquad Q_\pm = \frac12\left(
u^\dagger Q u \pm uQu^\dagger \right),
\ea%
where $u=\sqrt{U}$, $U= \exp \left( {\sqrt{2}i\phi / F_\pi}\right)$, with $F_\pi$
the pion decay constant, and $\phi$ collects the pseudoscalar mesons,
\begin{eqnarray}%
\label{eq:phi}
 \phi =
  \left(
    \begin{array}{c c c}
 \frac{1}{\sqrt{2}}\pi^0+\frac{1}{\sqrt{6}} \eta & \pi^+ & K^+ \\
\pi^- & - \frac{1}{\sqrt{2}}\pi^0+\frac{1}{\sqrt{6}} \eta & K^0 \\
K^- & \bar K ^0 & -\frac{2}{\sqrt{6}} \eta \\
    \end{array}
\right) .
\end{eqnarray}%
The light quark mass and charge matrices are
\ba
\label{eq:chiQ}
\chi = 2B_0\cdot {\rm diag}\left\{m_u,m_d,m_s\right\}, \quad Q =
e\cdot {\rm diag}\left\{\frac{2}{3},-\frac1{3},-\frac1{3}\right\}.
\ea

The light quark mass and em contributions to the heavy meson and doubly heavy
baryon masses can be easily worked out using this Lagrangian. At NLO, the
Goldstone bosons are not needed since the chiral loop corrections to the hadron
masses start from higher order. In this case, physically, the $d_1$ term
describes the effect of the virtual photons coupled to the light quark. Because
there is only one light quark in both heavy mesons and doubly heavy baryons,
these virtual photons merely contribute to the self-energy of the light quark.
So they can be absorbed into a redefinition of the quark masses
\ba%
\tilde{m}_u = m_u + \frac{d_1e^2}{9cB_0}, \qquad \tilde{m}_{d(s)} = m_{d(s)} +
\frac{d_1e^2}{36cB_0}.
\ea%
There is no $d_2$ term because $\lang Q_+\rang = \lang Q\rang$ vanishes with the
charge matrix given in Eq.~(\ref{eq:chiQ}). We remark that we are well aware of the
subtleties concerning  em corrections to quark masses (see
e.g. Ref.~\cite{Gasser:2003hk}) but these can be ignored to the accuracy we are working.

Therefore, at NLO there are only two parameters describing the mass corrections
of the heavy mesons and doubly heavy baryons. One is for the light quark mass
difference, and the other one is for the em effects which originate from the virtual
photons exchanged between the light quark and the heavy quarks. These two
effects can also be parameterized by two parameters in quark models once
neglecting the spin-dependent interactions which are suppressed by $1/m_Q$, see
e.g. \cite{Karliner:2008sv,Hwang:2008dj}. The explicit expressions for the mass
corrections can be found in the Appendix. The former parameter can be determined
through the measured SU(3) mass splitting between the heavy mesons, and the
latter one can be determined from the isospin mass splitting between the charged
and neutral heavy mesons~\cite{Guo:2008gp}. Defining $\tilde c\equiv
4cB_0(\tilde m_s-\tilde m_d)$ and $\tilde d\equiv d_3e^2/3$, using the mass
differences among the pseudoscalar charmed mesons
$D^0,D^+,D_s^+$~\cite{PDG2010}, one gets
\ba%
\tilde c = (98.99\pm0.30)~{\rm MeV}.
\ea%
The value of $\tilde d$ can be extracted via
\ba%
\tilde d = \frac12 [(M_{D_s^+}-M_{D^+})\lambda - (M_{D^+}-M_{D^0})] =
(-1.05\pm0.16)~{\rm MeV},
\ea%
where $\lambda=(\tilde m_d-\tilde m_u)/(\tilde m_s-\tilde m_d)=0.027\pm0.003$ is
calculated from the recent FLAG average~\cite{Colangelo:2010et}. This value for
$\lambda$ is consistent with the latest precise lattice determinations of the
light quark masses, $\lambda = 0.029\pm 0.002$~\cite{Durr:2010vn}. Note that in
both cases the value for $\lambda$ refers to the masses without the em shift.
In what follows, we will ignore this difference, as it is expected to be a minor
effect. Therefore, these values correspond to
\ba%
(M_{D^+}-M_{D^0})^{\rm strong} = (2.67\pm0.30)~{\rm MeV},\qquad
(M_{D^+}-M_{D^0})^{\rm em} = (2.10\pm0.32)~{\rm MeV},
\ea%
where the superscript ``strong'' and ``em'' denote the effects from the $u,d$
quark mass difference and the virtual photons, respectively. If we use the mass
differences among the pseudoscalar bottom mesons, $M_{B_s^0}-M_{B}  =
87.0\pm0.6$~MeV and $M_{B^0}-M_{B^+} = 0.33\pm0.06$~MeV~\cite{PDG2010}, then we
get different values
\ba%
\tilde c = (87.0\pm0.6)~{\rm MeV}, \qquad \tilde d = (-2.0\pm0.3)~{\rm MeV},
\ea%
which corresponds to the decomposition
\ba%
(M_{B^0}-M_{B^+})^{\rm strong} = (2.3\pm0.3)~{\rm MeV},\qquad (M_{B^0}-M_{B^+})^{\rm
em} = -(2.0\pm0.3)~{\rm MeV}.
\ea%
The values agree within uncertainties ---
the differences between these two sets of values may be understood as flavor
symmetry breaking corrections of order ${\cal O}(\Lambda_{\rm QCD}/m_c)$.

\begin{table}[t]
\begin{center}
\renewcommand{\arraystretch}{1.3}
\begin{tabular}{| l | c | c | c |} \hline
   & $M_{\Xi_{cc}^{++}}-M_{\Xi_{cc}^{+}}$  & $M_{\Xi_{bb}^{-}}-M_{\Xi_{bb}^{0}}$ & $M_{\Xi_{bc}^{+}}-M_{\Xi_{bc}^{0}}$ \\ \hline
EM & $\phantom{-}4.2\pm2.3$ & $4.0\pm1.5$ & $\phantom{-}1.6\pm1.1$\\
Strong & $-2.7\pm1.5$ & $2.3\pm0.8$ & $-2.5\pm1.4$ \\
Total & $\phantom{-}1.5\pm2.7$ & $6.3\pm1.7$ & $-0.9\pm1.8$ \\
\hline
Ref.~\cite{Hwang:2008dj} & $\phantom{-}2.3\pm1.7$ & $5.3\pm1.1$ & $-1.5\pm0.9$ \\
\hline
\end{tabular}
\caption{\label{tab:Bresults}Predicted isospin splitting of the doubly heavy
baryons. The electromagnetic and quark mass difference contributions are given
in the second and third rows, respectively. The final results are shown in the
fourth row. The results in Ref.~\cite{Hwang:2008dj} are given in the last row
for comparison. All values are given in units of MeV.}
\end{center}
\end{table}
Using these parameter values, the mass difference between the $\Xi_{cc}^{++}$
and $\Xi_{cc}^{+}$ can be easily obtained. One can also get the isospin
splittings for the doubly bottom baryons $\Xi_{bb}$ and the bottom-charm baryons
$\Xi_{bc}$. All of the predictions are listed in Table~\ref{tab:Bresults}. To
minimize the uncertainty from heavy quark flavor symmetry, the results for the
$\Xi_{cc}$ and $\Xi_{bb}$ are given using the values of $\tilde c$ and $\tilde
d$ extracted from the charmed and bottom mesons, respectively. The results for
the $\Xi_{bc}$ cover the values obtained using both sets of parameter values.
Now let us discuss the other possible uncertainties. Except for the ones from
$\tilde c$ and $\tilde d$, there are still uncertainties from neglecting higher
order counterterms and loops in the chiral expansion. Since our predictions
concentrate on the isospin splittings, the relevant corrections from higher
order terms in the chiral expansion should be of order ${\cal
O}(M_\pi/\Lambda_\chi)\sim 15\%$, with $M_\pi$  the pion mass and
$\Lambda_\chi\simeq1$~GeV the chiral symmetry breaking scale. In addition, there
should also be corrections to the heavy quark-diquark U(5) symmetry. These
corrections should be of order ${\cal O}(r_{QQ}/r_{qQ})={\cal O}(\Lambda_{\rm
QCD}/(m_Qv))$ which describes the relative size of the neglected heavy quark
separation with respect to the distance between the light quark and the heavy
diquark. Conservatively, we take ${\cal O}(\Lambda_{\rm QCD}/(m_Qv))\sim
50\%,30\%$ and $50\%$ for the $\Xi_{cc}$, $\Xi_{bb}$ and $\Xi_{bc}$ baryons,
respectively. The same isospin splittings were also calculated in an approach
based on a parameterization inspired by heavy quark
effective theory and utilizing some data to fix the parameters~\cite{Hwang:2008dj}. For comparison, their
results are given in the last row.

One can get both spin-3/2 and 1/2 doubly heavy baryons from binding a spin-1
heavy diquark and a light quark. Because the spin of the diquark is the same
in both cases, they are related to each other by the heavy quark spin
symmetry, and have the same isospin splittings as given in
Table~\ref{tab:Bresults}.  Corrections to the spin symmetry are suppressed by
$\Lambda_{\rm QCD}/m_Q$.  These corrections are expected to be small as
confirmed by a comparison with the results of a quark
model~\cite{Karliner:2008sv} which takes into account the spin-dependent
interactions, namely $M_{\Xi_{bb}^{-}}-M_{\Xi_{bb}^{0}}=6.24\pm0.21$~MeV or
$6.4\pm1.6$~MeV using different inputs, which is quite close to ours.

\section{Isospin splittings of quadratically heavy pentaquarks}
\label{sec:pentaquark}

The analysis can be extended to pentaquarks containing four heavy quarks and one
light quark. One should notice that the size of the four heavy quark cluster is
not four times larger than the distance between two heavy quarks, $r_{QQ}$.
Being in an $S$-wave, if the four quarks are of the same flavor, they should be
spatially symmetric. Hence, the size of the cluster is the same as $r_{QQ}\sim
1/(m_Qv)$, which is again much smaller than $1/\Lambda_{\rm QCD}$ in the heavy
quark limit. Moreover, in a pentaquark, the four heavy quarks are in a
fundamental representation of the SU(3) color symmetry group, which is the same as
one quark. Hence, to a first approximation they can be treated as one object, to
be called quadra-quark in the following. For the four heavy quarks of the same
flavor being in an $S$-wave, Fermi statistics requires their spin wave function
to be symmetric. Hence, the quadra-quark is a spin-2 object. Analogous to the
U(5) symmetry between the heavy diquark and heavy anti-quark, the symmetry for
the heavy quadra-quark and heavy quark is U(7). One may find an interesting
phenomenology for the pentaquarks using the U(7) symmetry. Here, we are only
interested in the isospin splittings. It is easy to find the following relations
to lowest order in isospin breaking:
\ba%
M_{cccc\bar d} - M_{cccc\bar u} \al=\al 4 (M_{D^+}-M_{D^0})^{\rm em} +
(M_{D^+}-M_{D^0})^{\rm strong} = (11\pm5)~{\rm MeV}, \non\\
M_{bbbb\bar d} - M_{bbbb\bar u} \al=\al 4 (M_{B^0}-M_{B^+})^{\rm em} +
(M_{B^0}-M_{B^+})^{\rm strong} = -(6\pm3)~{\rm MeV}.
\ea%
The splittings are large, but certainly more of an intellectual curiosity at present.

\section{Summary}

Using the Cottingham formula to compute the Coulomb electromagnetic shift, we
have shown that the large SELEX value of the isospin splitting of the $\Xi_{cc}$
states implies that double charm baryons are very compact; i.e., the light quark
must be very close to the two heavy quarks.  A novel possibility is that the
quarks in the doubly charm baryons are arranged as a compact state  $c-q-c$ with
a linear geometry.  This possibility could be checked using lattice gauge theory
simulations. However, the infrared behavior of the light quark is expected to be
governed by the non-perturbative confining interaction, and thus the size of any
hadron containing a light quark should be of order ${\cal O}(1/\Lambda_{\rm
QCD})$. A conventional approach exploiting this is based on quark-diquark
symmetry. It  allows us to
predict the isospin splitting for doubly heavy baryons $\Xi_{cc}$, $\Xi_{bb}$
and $\Xi_{bc}$  at NLO in the chiral expansion. These predictions for the doubly
charm baryons give isospin separations much smaller than the SELEX measurements.
Therefore, the compactness implied by the SELEX data appears to
call for a significant violation of heavy quark--diquark symmetry --- today no
mechanism is known that can provide this.
 However, it should be noticed that among all the four $\Xi_{cc}$
states in the two reported isospin doublets only the mass of the
$\Xi_{cc}^+(3520)$ has been measured with certainty.

In order to resolve the discrepancy between the experiment and theory, further
precise experimental investigations are clearly needed. Our prediction for the
$\Xi_{bb}$ isospin splitting is $6.3\pm1.7$~MeV based on heavy quark diquark
symmetry. It is of similar size as $M_{\Xi^-}-M_{\Xi^0}$ which contain two
strange quarks. (Because $m_s<\Lambda_{\rm QCD}$, we have refrained from making
use of the same method to calculate the isospin splittings containing strange
quark(s).)  Any configuration that leads to isospin splittings as large as those
reported by SELEX for the doubly charmed baryons would lead to significantly
larger splittings for the doubly bottomed baryons, because the em self-energy
would be much larger.

 We have also made predictions for the isospin splittings of the pentaquarks
 $cccc\bar q$ and $bbbb\bar q$ and have found that the value for the $cccc\bar
 q$ is as large as $(11\pm5)$~MeV using a generalization of heavy quark
 diquark symmetry.

 As we have noted, the best chance to create these
 super-heavy hadrons and test their properties is in hadron-hadron collisions
 at high $x_F$ using the intrinsic heavy quark Fock state mechanism --- for
 example, at the LHCb, or at future fixed-target experiments using the 7~TeV
 LHC beam.

\medskip

\section*{Acknowledgments}
We want to thank J.~Gasser for helpful discussions and U.-G.M. acknowledges a
useful communication from H. Leutwyler.  We also thank J. Engelfried, A.
Goldhaber, M. Karliner, H. Lipkin and J. Russ for comments. F.-K.G., C.H. and
U.-G.M. would like to thank  the DFG (SFB/TR 16, ``Subnuclear Structure of
Matter''), the HGF (Virtual Institute ``Spin and Strong QCD'', VH-VI-231), and
the European Community-Research Infrastructure Integrating Activity ``Study of
Strongly Interacting Matter'' (acronym HadronPhysics2, Grant Agreement n.
227431) under the Seventh Framework Programme of EU for support. U.-G.M. also
thanks the BMBF for support (grant 06BN9006). This research was also supported
by the Department of Energy, contract DE--AC02--76SF00515.

\medskip

\begin{appendix}

\section{Expressions for the mass corrections at NLO}
\label{app:LO}
\renewcommand{\theequation}{\thesection.\arabic{equation}}
\setcounter{equation}{0}

In the appendix, we give explicit expressions for the NLO corrections of the
masses of the heavy mesons and doubly heavy baryons. In the heavy quark limit,
the QCD Lagrangian does not depend on the heavy quark mass, which results in the
heavy quark flavor symmetry, see e.g.~\cite{Neubert:1993mb}. In the following,
the formula are given in terms of general heavy quark flavors,
\ba%
Q\bar u: \al\al 4cB_0\tilde m_u + \frac{2}{3}e_Qd_0e^2, \qquad
Q_1Q_2u:~  4cB_0\tilde m_u - \frac{2}{3}(e_{Q_1}+e_{Q_2})d_0e^2, \non\\
Q\bar d: \al\al 4cB_0\tilde m_d - \frac1{3}e_Qd_0e^2, \qquad
Q_1Q_2d:~  4cB_0\tilde m_d + \frac1{3}(e_{Q_1}+e_{Q_2})d_0e^2, \non\\
Q\bar s: \al\al 4cB_0\tilde m_s - \frac1{3}e_Qd_0e^2, \qquad %
Q_1Q_2s:~ 4cB_0\tilde m_s + \frac1{3}(e_{Q_1}+e_{Q_2})d_0e^2,
\ea%
where $Q_{(1,2)}$ represents the heavy quark flavor, and $e_{Q_{(1,2)}}$ its
charge in unit of the elementary charge $e$ with $e>0$.

\end{appendix}

\end{document}